\documentclass[12pt,a4paper]{article}
\pdfoutput=1
\usepackage[utf8x]{inputenc}

\usepackage{amssymb}
\usepackage{amsmath}
\usepackage{graphicx}
\usepackage{hyperref}

\newcommand{\eq}{\begin{equation}}
\newcommand{\eqx}{\end{equation}}
\newcommand{\eqs}{\begin{equation*}}
\newcommand{\eqsx}{\end{equation*}}
\newcommand{\eqn}{\begin{eqnarray}}
\newcommand{\eqnx}{\end{eqnarray}}
\newcommand{\eqns}{\begin{eqnarray*}}
\newcommand{\eqnsx}{\end{eqnarray*}}
\newcommand{\alg}{\begin{align}}
\newcommand{\algx}{\end{align}}

\newcommand{\f}[2]{\frac{#1}{#2}}

\newcommand{\lm}{\lambda}

\newcommand{\sg}{\sigma}

\newcommand{\dl}{\delta}
\newcommand{\Dl}{\Delta}
\newcommand{\al}{\alpha}
\newcommand{\bt}{\beta}
\newcommand{\om}{\omega}

\newcommand{\gm}{\gamma}
\newcommand{\Gm}{\Gamma}

\newcommand{\qq}{\quad\quad}

\newcommand{\tr}{\mbox{\rm tr}\,}

\newcommand{\nn}{{\cal N}}

\newcommand{\slii}{{\tt sl(2)}\ }
\newcommand{\Qt}{\tilde{Q}}
\newcommand{\oo}[1]{{\mathcal O}\left(#1\right)}

\title{Twist-two operators and the BFKL regime\\
-- nonstandard solutions of the Baxter equation}

\author{Romuald A. Janik\thanks{e-mail: {\tt romuald@th.if.uj.edu.pl}}}

\date{\vspace{6pt}Institute of Physics, Jagiellonian University\\
ul. Reymonta 4, 30-059 Kraków, Poland\\}

\begin{document}

\maketitle

\begin{abstract}
The link between BFKL physics and twist-two operators involves
an analytical continuation in the spin of the operators
away from the physical even integer values. Typically this
is done only after obtaining an analytical result for integer spin
through nested harmonic sums.
In this paper we propose analyticity conditions for the
solution of Baxter equation which would work directly
for any value of complex spin and reproduce results
from the analytical continuation of harmonic sums. 
We carry out explicit contructions up to 2-loop level.
These nonstandard
solutions of the Baxter equation have rather surprising
asymptotics. We hope that these analyticity conditions may be used
for incorporating them into the exact TBA/FiNLIE/QSC approaches
valid at any coupling.
\end{abstract}

\vfill{}

\section{Introduction}

A very important dynamical regime of gauge theory is the Regge limit
of high energy scattering characterized by very high energy and fixed momentum
transfer (equivalently this corresponds to the small $x$ regime of Deep Inelastic
Scattering in QCD). In this regime, scattering amplitudes behave as a power of the energy.
A perturbative description at leading order involves a resummation
of all terms contributing as $\lm \log s$ and yields the (LO) BFKL pomeron 
\cite{BFKL} (and its generalizations to states with more than two reggeized gluons). Currently we know also results at the NLO level both in QCD and in $\nn=4$ SYM
\cite{NLO}. However, there does not seem to be any chance of going directly
beyond NLO using only standard perturbative computations.

In the context of $\nn=4$ SYM theory, further progress can be achieved using the
methods of the AdS/CFT correspondence. 
At strong coupling the scattering amplitudes behave like $s^2$ (thus the strong
coupling pomeron intercept is 2) with the leading contribution
coming from graviton exchange \cite{JP}. Subsequently, the first correction in $1/\sqrt{\lm}$
was determined in \cite{BPS}. Recently, significant progress was made
due to the link with twist-2 operators. 
Indeed, currently we know 3 further terms in the strong coupling expansion of
the intercept \cite{COSTA,LipLast}.

BFKL physics is very interesting and important for a number of reasons.
Firstly, the pomeron intercept is an example of an IR safe observable relevant
for high energy scattering. Secondly, LO BFKL is exactly the same in QCD and
in $\nn=4$ SYM. At the NLO level, differences appear, but it would be very interesting
to understand completely the conformal physics of the pomeron. On a more theoretical
side, the multi-reggeized gluon dynamics at the LO BFKL level in QCD was probably the
first place were integrability was discovered in a four dimensional gauge theory. However
even now we do not know if, and in what sense, is NLO BFKL integrable.
Another fascinating issue is the question how does BFKL fit into our
very complete understanding of integrability of the spectral problem in $\nn=4$ SYM.
This has to be a very nontrivial link as it is known that even LO BFKL involves
an infinite set of wrapping corrections, so any relation between BFKL and AdS/CFT
integrability will have to be on the full $AdS_5 \times S^5$ $\sg$-model level.
Even all-loop Asymptotic Bethe Ansatz will not suffice.

Taking all the above into account, 
it is not clear, however, what is the optimal approach to the study of all-order BFKL physics
from the point of view of AdS/CFT integrability. One could either attempt a direct
approach dealing with observables directly linked to the pomeron, or a more indirect
approach which uses the very close links between BFKL and the anomalous dimensions of
twist-2 operators about which we have currently quite detailed knowledge.
In this paper we will pursue this latter approach, leaving the more direct approach
to a forthcoming paper \cite{TOAPPEAR}. 

The key relation which links the anomalous dimensions of twist-2 operators
and the pomeron intercept involves the analytical continuation of
these anomalous dimensions away from the physical values of even integer spin $M$.
This procedure, which is technically quite demanding, involves finding
first an \emph{analytical} expression for the anomalous dimensions
as a function of the spin in terms of so-called nested harmonic sums.
Then, one has to find an appropriate analytical continuation of the harmonic
sums to arbitrary values of the spin $M$ (and, at weak coupling, analyze the
pole structure at $M=-1$).

This procedure, although very involved, has been carried out including Bethe
ansatz results \cite{KLRSV} and wrapping corrections at 4- \cite{TWISTTWO4} and
5-loop level~\cite{TWISTTWO5}. However, once we would want to study the problem
at finite coupling, where we have mainly numerical approaches like TBA (Thermodynamic Bethe
Ansatz) \cite{TBA}, FiNLIE (Finite Nonlinear Integral Equations) \cite{FINLIE,BHNLIE}
(and currently the Quantum Spectral Curve (QSC) \cite{QSC}), this approach is doomed
to failure since we cannot perform an analytical continuation from numerical
data at integer points. 

The motivation for this work is to develop methods for working directly at
any complex values of the spin $M$ in a way which is compatible with the
known analytical continuations of the nested harmonic sums.
Since the basic building block of the spectral problem is
a Baxter function (in this context a solution of Baxter equations in the \slii sector
which should be generalized to the whole T/Y-system once we include
arbitrary wrapping corrections), we will propose certain analyticity conditions for
the behaviour of the Baxter function for any complex $M$ which 
would reproduce the analytical continuations of harmonic sums at 1- and 2-loop
level. This is the main result of the present paper.
We expect, although we do not have a proof, that these analytical
conditions should be valid in much more generality. We hope that they can be used in order to 
formulate a TBA/FiNLIE/QSC problem directly for any complex $M$.
Solving these equations would then potentially provide information about
BFKL physics valid at any coupling.

The plan of this paper is as follows. In section 2 we will give a brief introduction
to the anomalous dimensions of twist-2 operators in the \slii sector and
state more explicitly their link with BFKL. In section 3 we will review the main properties
of 1-loop Baxter equation and in section 4 we will formulate our key proposal for the analytical
properties of the physical solution of the Baxter equation at any (non-integer) value
of the spin. In section 5 we will explicitly construct the relevant solution
of the 1-loop Baxter equation and perform various checks. In section 6 we will
show how to extend this solution to the 2-loop level. We close the paper
with a summary and outlook and several appendices with some technical details. 

\section{Twist-two operators and harmonic sums}

Twist-two operators in the \slii sector are formed out of two complex scalar fields
and an arbitrary number $M$ of derivatives ($\equiv$ spin of the operator) along 
a fixed light-cone direction.
\eq
O_M= \tr Z D_+^M Z+ \ldots
\eqx
For each even integer $M$, there appears one new primary state, and its dimension
defines the function $\Dl(M)$ for even integer $M$'s.
It is known, currently up to 5-loop level \cite{KLRSV,TWISTTWO4,TWISTTWO5}, that $\Dl(M)$ 
is expressed in terms of
nested harmonic sums. For example up to 2-loop level we have the expression
\eq
\Dl(M)= 2+M+\gm(M) \equiv 2+M+ g^2 \cdot \gm^{(1-loop)}(M)+ g^4 \cdot \gm^{(2-loop)}(M)+\ldots
\eqx
where
\eqns
\!\!\!\! \gm^{(1-loop)}(M)\!\!\!\! &=&\!\!\!\!  8\, S_1(M) \\
\!\!\!\! \gm^{(2-loop)}(M)\!\!\!\!  &=&\!\!\!\!  -16(S_3(M)+S_{-3}(M)-2 S_{-2,1}(M)+2 S_1(M) (S_2(M)+S_{-2}(M)))
\eqnsx
These expressions obey the maximal transcendentality principle, which up to now still
remains mysterious, which states that the degree of transcendentality\footnote{Defined
as the sum of absolute values of the harmonic sum indices and the arguments of $\zeta$
values if they appear.} of all terms is maximal and equal to $2n-1$, where $n$ is the loop order
of the perturbative computation.
Even more mysteriously the maximal transcendentality part of the QCD answer
exactly coincides with the above expressions for $\nn=4$ SYM.

The harmonic sums are defined as
\eq
S_k(M)=\sum_{j=1}^M \f{1}{j^k} \qq
S_{k,l}=\sum_{j=1}^M \f{1}{j^k} S_l(j)
\eqx
for positive indices, and 
\eq
S_{-k}(M)=\sum_{j=1}^M \f{(-1)^k}{j^k} \qq
S_{-k,l}=\sum_{j=1}^M \f{(-1)^k}{j^k} S_l(j)
\eqx
for negative (or mixed) indices.
These functions have a well defined analytical continuation (such that the only
singularities appear on the negative real axis)
to \emph{complex} values of $M$ \cite{ANALCONT} e.g.
\eq
S_1(M) = \Psi(1+M)-\Psi(1)
\eqx
but which becomes rapidly more complicated for nested sums 
and especially for sums with some negative indices (c.f. \cite{ANALCONT}). 
In appendix~\ref{s.analcont}, we give formulas for the specific harmonic sums that 
we will use in this paper.

Of particular interest to this paper and its primary motivation is the
relation between the anomalous dimensions of twist-2 operators 
analytically continued to $M=-1+\om$ and the BFKL pomeron intercept $j(\gm)$.
Indeed the singularities of the anomalous dimension $\gm$ as a function of
$\om$ can be extracted from the BFKL pomeron intercept through the
relation
\eq
\label{e.omjgm}
\om=j(\gm)-1
\eqx
The relation between anomalous dimensions and BFKL was first proposed by Jaroszewicz 
\cite{Jaroszewicz}, and exploited in \cite{NLO}. In the $\nn=4$ SYM integrability
context it was used by \cite{KLRSV} to show explicitly the neccessity of wrapping 
corrections through a contradiction between Bethe ansatz results at 4-loop level
and the BFKL predictions from (\ref{e.omjgm}). 
The inclusion of wrapping corrections, first at 4-loop \cite{TWISTTWO4} and then
at 5-loop level \cite{TWISTTWO5} resolved this contradiction. 
The above relation between BFKL and twist-2 operators show the neccessity of 
computing the analytical continuations of the anomalous dimensions of these operators 
for generic \emph{noninteger} values of the spin $M$. The
investigation of this issue is the main focus of the present paper.

\section{Baxter equation in the \slii sector}

The anomalous dimensions of twist-two operators with even integer spin $M$ 
can be described using Bethe ansatz with $M$ Bethe roots corresponding to 
the $M$ excitations, each carrying a unit of spin. As mentioned in the introduction,
it is not possible to describe the analytical continuation of these states
to noninteger (generally complex) $M$ within this framework.
In this paper we will therefore use a standard reformulation of the Bethe ansatz
in terms of the Baxter equation, which in fact holds for arbitrary complex~$M$.

In the simplest case of 1-loop anomalous dimensions, the Bethe ansatz equations
read
\eq
\label{e.bae1}
\left( \f{u_j+\f{i}{2}}{u_j-\f{i}{2}} \right)^2= \prod_{\substack{k=1\\k\neq j}}^M \f{u_k-u_j-i}{u_k-u_j+i}
\eqx  
while the corresponding Baxter equation takes the form
\eq
\label{e.baxter1}
\left(u+i/2\right)^2 Q(u+i) +\left(u-i/2\right)^2 Q(u-i) =   \underbrace{\left(2u^2-M(M+1) 
-\f{1}{2} \right)}_{t^{1-loop}(u;M)} Q(u)
\eqx
If $M$ is integer, then a polynomial solution of (\ref{e.baxter1}) is equivalent
to (\ref{e.bae1}) with the zeroes of the polynomial $Q(u)$ being identified with
the Bethe roots appearing in (\ref{e.bae1}).

Once we relax the condition of integrality of $M$, we have to determine how to 
pick the \emph{physical solution} which would coincide with the standard analytical continuation
of harmonic sums which determine the energies (anomalous dimensions) and all higher
conserved charges of the twist-two states.
In the following section we will formulate our proposal for the analyticity conditions
which would single out the physical solution for any complex $M$.

In this section we will briefly review some standard properties of the (1-loop) Baxter
equation (\ref{e.baxter1}). It is clear that the solutions of Baxter equation are determined
up to multiplication by an overall periodic function:
\eq
Q(u) \to f(u) Q(u) \qq \text{with} \qq f(u+i)=f(u)
\eqx
This is just a gauge symmetry without any physical consequences.
A convenient way to factor it out is to introduce the ratio \cite{Basso}
\eq
R(u)=\f{Q(u+i/2)}{Q(u-i/2)}
\eqx
$R(u)$ also summarizes two infinite sets of conserved charges appearing in its
expansion around $u=0$ and $u=\infty$
\eq
\log R(u) =\sum_{n=1}^\infty i \f{Q_{-n}}{u^n}  \qq \log R(u) =\sum_{n=1}^\infty i Q_n u^n
\eqx
Of particular interest for us will be the family of conserved charges with positive indices 
$Q_{n>0}$ (which include in particular the energy). They are expressed as polynomials
of $Q(u)$ and its derivatives evaluated at the special points $u=\pm i/2$.
It is convenient to use the normalization $Q(i/2)=1$.

From \emph{polynomial} solutions of the Baxter equation for even $M$, there
are explicit expressions for these derivatives (up to a few first orders) in
terms of harmonic sums \cite{KRZ,BBKZ}. In particular we have\footnote{Here we have used
some identities between harmonic sums given in \cite{HSrelations}. See appendix~\ref{s.analcont}.}
\eqn
\label{e.Qders1}
Q(i/2) &=& 1 \qq \text{(normalization)}\\
Q'(i/2) &=& -2i\, S_1 \\
Q''(i/2) &=& -4(S_{-2}+S_1^2) \\
Q'''(i/2) &=& -24i \left( -2 S_{-2,1}-S_1 S_{-2}-S_{-3}+\f{1}{3} S_3-\f{1}{3} S_1^3 \right)
\label{e.Qders2}
\eqnx
The first derivative is just equivalent to the 1-loop energy formula $E=\gm^{(1-loop)}(M)=8\,S_1(M)$.
From the above expressions we see that the degree of transcendentality
is equal to the order of the derivative. 
Once we replace the harmonic sums by their standard analytical continuation,
we will want to reproduce the above expressions directly from our solution
at noninteger $M$.

In addition we have also a closed form expression for all charges $Q_{n>0}$ up
to linear order in $M$ \cite{Basso,G} which at 1-loop can be conveniently
expressed as
\eq
\log R(u) = \f{i M}{u}-\f{i M \sinh(2\pi u)}{2\pi u^2} +\oo{M^2}
\eqx 
This will be again an important cross-check of our solution.

The Baxter equation is a second order difference equation and thus one expects
two linearly independent solutions. In fact, once we have a generic (neither even or odd) 
solution $\Qt_1(u)$, the second solution can be taken to be $\Qt_1(-u)$. 
However in contrast to the case of second order differential equations, the total space
of nonequivalent solutions is in fact infinite dimensional
\eq
\label{e.Qtwosol}
Q(u)=f_1(u,M) \Qt_1(u) + f_2(u,M) \Qt_1(-u)
\eqx
with $f_1(u,M)$ and $f_2(u,M)$ being arbitrary periodic functions with period $i$.

Let us note that there is a well known solution of the Baxter equation
valid for arbitrary complex $M$  \cite{KorQt}:
\eq
\Qt_1(u)={}_3F_2(-M,M+1,1/2-iu;1,1|1)
\eqx 
In fact this solution reduces to the correct polynomial solution for integer $M$.
However it is \emph{not} the correct physical analytical continuation as
it has an explicit $M \to -1-M$ symmetry, which is not a symmetry of
the anomalous dimensions $\gm^{(1-loop)}=8\,S_1(M)$. Nevertheless $\Qt_1(u)$ turns out
to be a convenient building block of the physical solution as in (\ref{e.Qtwosol}). 
We will often refer to it as \emph{the elementary solution} in the following. 

Finally, by expanding the Baxter equation at large $u$, one can see that
there are two possible large $u$ asymptotics of its solutions:
\eq
Q(u) \sim u^M \qq \text{or} \qq Q(u) \sim u^{-1-M}
\eqx
This leads to
\eq
\log R(u) \sim \f{i M}{u}+ \oo{1/u^2}  \qq \text{or} \qq \log R(u) \sim \f{-i (1+M)}{u}+ \oo{1/u^2}
\eqx
The first choice reduces to the well known polynomial solutions for integer $M$. Surprisingly,
it turns out that it is the other choice which singles out the physical solution
for complex $M$.

\section{The key proposal} 
\label{s.key}

We will now formulate a proposal on the analytical conditions that should single
out a particular \emph{`physical'} solution of the Baxter equation for any complex $M$,
such that all physical properties, like anomalous dimensions and higher conserved
charges, computed from this solution would coincide with the ones obtained from
standard analytical continuation of harmonic sums appearing in the expressions
for even integer $M$. We will then proceed to test the above proposal 
at the 1- and 2-loop level. 

{\bf Claim:} The solution which reproduces all known
constraints (BFKL, small $M$ charges) is an even entire function (i.e. with no poles) characterized by the asymptotics\footnote{At higher loop orders, the $u^{-M-1}$ 
asymptotics will get modified by logarithmic terms.}
\eq
\log R(u) \sim \f{-i (1+M)}{u}+ \oo{1/u^2}
\eqx
at $u \to \pm \infty$. More precisely, the component of $Q(u)$ with asymptotics $\sim u^M$
(which is not modified at higher loop orders)
should vanish (or be exponentially suppressed\footnote{E.g. like $u^{-M-1}+u^M/\sinh(2\pi u)$.}) at $u \to \pm \infty$.

Let us note that this proposal is extremely counterintuitive and unexpected.  
The above asymptotic condition is in direct contradiction with the physical polynomial
solutions at even $M$ which behave at infinity like $u^M$. And it is just from these
solutions that we get all our information about the energies and charges\footnote{The $u^M$
asymptotics was even proposed away from integer $M$ in \cite{Basso}. We will, however,
recover the correct small $M$ values of the charges also from our solution.}. It will turn
out, however, that the relation between the polynomial solutions and our complex $M$
solution is quite subtle. We will discuss this point at the end of the following section.

\section{The 1-loop solution}

In this section we will construct the solution of the 1-loop Baxter equation which
satisfies the analiticity requirements spelled out in section \ref{s.key}.
First we will analyze the pole structure of the elementary solutions and then
we will impose the appropriate asymptotics at infinity.

In order to study the properties of the elementary solution $\Qt_1(u)$ it is necessary to
have a convergent representation of this function for any $u$. It turns out that the
standard power series representation of hypergeometric functions applied to $\Qt_1(u)$
yields 
\eq
\label{e.lower1}
\Qt_1(u)= \f{1}{\Gm(-M) \Gm(M+1) \Gm(1/2-iu)} \sum_{k=0}^\infty \f{\Gm(k-M) \Gm(k+M+1)
\Gm(1/2-i u+k)}{k!^3}
\eqx
and gives a valid representation only for $Im\, u<0$. An alternative expression which is valid for 
$Im\, u>0$ can be obtained using the results and methods of \cite{LipatovDV}
\eqn
\label{e.upper1}
\Qt_1(u)&=& \f{\cosh \pi u}{\pi} \f{\sin^2 M\pi}{\pi^2} \Gm(1/2-iu) 
\sum_{k=0}^\infty \f{\Gm(k-M) \Gm(k+M+1)
\Gm(1/2+i u+k)}{k!^3} \cdot \nonumber\\
&& \cdot \left( 3\psi(k+1)-\psi(k-M)- \psi(k+M+1)-\psi(1/2+i u+k) \right)
\eqnx

Using the above representations, we may derive the behaviour of the elementary solution
$\Qt_1(u)$ at $u=\pm i/2$ as these points are crucial for the determination of the energy and
higher conserved charges.
We find the following behaviour of $\Qt_1(u)$ at $u\sim i/2$:
\eq
\Qt_1(u) \sim \f{i \sin M\pi}{\pi (u-i/2)} +\left( \cos M\pi +\f{2}{\pi} \sin(M \pi) S_1(M) \right) + \oo{u-i/2}
\eqx
and at $u\sim -i/2$:
\eq
\Qt_1(u) \sim \sim 1 + \oo{u+i/2}
\eqx
We see that once we move away from integer $M$, a pole appears at $u=i/2$. In order to deal
with entire functions we will cancel the poles by multiplying the elementary solutions
by an overall periodic function
\eq
\sinh(2\pi u) \Qt_1(u) \qq \text{and} \qq \sinh(2\pi u) \Qt_1(-u)
\eqx

One can convince oneself (see appendix~\ref{s.appQt}) that the asymptotics of $\Qt_1(u)$ at $u \to +\infty$ are 
\eq
\label{e.qtiasympt}
\Qt_1(u) \sim e^{i\pi \f{M}{2}} \f{\Gamma(1+2M)}{\Gamma(1+M)^3} u^M \left(1+\ldots\right)
-i e^{-i\pi \f{M}{2}} \f{\Gamma(-1-2M)}{\Gamma(-M)^3} u^{-1-M} \left(1+\ldots\right)
\eqx
Asymptotics at $u \to -\infty$ follow by complex conjugation as $\Qt_1(-u)=\Qt_1(u)^*$.
Hence, if we would like to cancel the $u^M$ component in the asymptotics at $u\to \pm \infty$,
our solution should reduce to the following combinations of the elementary solutions (up
to an overall factor of $\sinh 2\pi u$):
\eqn
u \to +\infty  &\qq& e^{-i\pi \f{M}{2}} \Qt_1(u) - e^{i\pi \f{M}{2}} \Qt_1(-u) \\
u \to -\infty  &\qq& e^{i\pi \f{M}{2}} \Qt_1(u) - e^{-i\pi \f{M}{2}} \Qt_1(-u)
\eqnx 
This shows that the periodic functions appearing in (\ref{e.Qtwosol}) are indeed
nontrivial. We will constrain them by the requirement that these functions should not introduce 
any poles into the solution. An essentially unique minimal choice can be constructed
out of constants and $\coth \pi u$ ($\tanh\pi u$ would introduce poles at $u=\pm i/2$, while
the poles of $\coth \pi u$ are canceled by the overall factor $\sinh 2\pi u$ which is
already in place).

The final 1-loop solution $Q_1(u)$ is consequently given by\footnote{The same solution
was found independently by N. Gromov and V. Kazakov \cite{GKpriv}}
\eq
\label{e.Qone}
\!\!\!\!\!\f{ \f{1}{2} i \sinh (2\pi u)}{\sin \pi M} \left[
\left(1\!-\!i\,\tan \f{\pi M}{2} \coth \pi u\right) \Qt_1(u) -
\left(1\!+\!i\,\tan \f{\pi M}{2} \coth \pi u\right) \Qt_1(-u) \right]
\eqx

Let us comment on some features of the above solution. The overall $M$ dependent factor
ensures the normalization
\eq
Q_1(i/2)=1
\eqx
which is convenient for comparision with the formulas (\ref{e.Qders1})-(\ref{e.Qders2}).
The solution is even in $u$ for any complex $M$. We checked analytically that $Q_1(u)$
reproduces the correct 1-loop energy\footnote{This is equivalent to showing that 
$Q'(i/2)=-2i\, S_1(M)$}
\eq
\gm^{(1-loop)}(M)=8\, S_1(M)
\eqx
as well as it reproduces all charges at the linear level in $M$:
\eq
\log R(u) = \f{i M}{u}-\f{i M \sinh(2\pi u)}{2\pi u^2} +\oo{M^2}
\eqx
Furthermore we checked numerically for sample values of $M=-9/10,\, 2/3$ and  $33/10$
that the formulas (\ref{e.Qders1})-(\ref{e.Qders2}) are satisfied to a very high precision.
This is a very nontrivial test as e.g. $Q'''(i/2)$ involves nested harmonic sums
of transcendentality 3, some with negative indices, whose analytical continuation
is quite intricate (see appendix~\ref{s.analcont} for details). 

Let us point out that the above mentioned tests were indeed necessary. In fact it is not
enough to require only that the 1-loop energy $\gm^{(1-loop)}(M)=8\, S_1(M)$ is recovered
in order to single out a unique solution of the Baxter equation. Indeed one can check
analytically that the \emph{two} apparently much simpler solutions
\eq
\f{\Qt_1(u)-e^{\pm i M \pi} \Qt_1(-u)}{\sin M \pi}
\eqx
also give $\gm^{(1-loop)}(M)=8\, S_1(M)$. However they do not agree with what we know
about the higher charges in the small $M$ limit. Indeed they have nonzero odd charges
and in fact correspond to the two families of charges $q_n^\pm$ defined in \cite{Basso}
--- while the physical solution has charges $\f{1}{2}(q_n^+  + q_n^-)$.

\begin{figure}[t]
\centerline{\includegraphics[height=7cm]{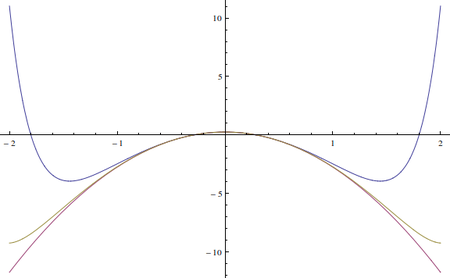}}
\caption{The polynomial solution for $M=2$ (red, bottom curve) and $Q_1(u)$ for $M=2.1$ (blue, top curve) and $M=2.01$ (yellow, middle curve)}
\end{figure}

Finally it remains to discuss the case of integer spin. Then the requirement for our solution,
namely its $u^{-M-1}$ asymptotics is in apparent outright contradiction with what we know
on the physical polynomial solutions, which in fact were used to derive formulas
(\ref{e.Qders1})-(\ref{e.Qders2}). How is this possible?

Let us note that the solution $Q_1(u)$ defined in (\ref{e.Qone}) behaves in a very nontrivial
way when $M$ approaches an integer. In figure~1 we show $Q_1(u)$ evaluated numerically  
for $M=2.1$ and
$M=2.01$ and compared it with the standard polynomial solution for $M=2$. We see that
the \emph{pointwise} limit of $Q_1(u)$ when $M \to 2$ is the standard polynomial solution.
However the limits $u \to \pm \infty$ and $M \to 2$ do not commute and this explains the
apparent contradiction. In fact, despite appearances, the solution $Q_1(u)$ is \emph{not} related to the
second nonpolynomial solution of the Baxter equation when $M$ is an integer.

\section{The 2-loop solution}


In this section we will show how to extend the 1-loop solution obtained in
the previous section to the 2-loop level.

Here we face a couple of technical difficulties. Although the 2-loop analog of the
$\Qt_1(u)$ solution is known explicitly, it is much more difficult to study its
analytical properties. Moreover, it will turn out that $\Qt_2(u)$ is more singular at
the crucial points $u=\pm i/2$ and we will have to supplant the solution by
appropriate choice of $g^2$ times a 1-loop solution in order to cancel the poles.

First we will discuss the modifications of the Baxter equation at two loops and then the
construction of the solution.

Let us write the Baxter function up to 2-loops as
\eq
Q(u)=Q_1(u)+g^2\, Q_2(u)+ \oo{g^4}
\eqx
Then the Baxter equation for $Q_2(u)$ is
\eqn
&&\!\!\!\!\!\!\!\left(u+i/2\right)^2 Q_2(u+i) +\left(u-i/2\right)^2 Q_2(u-i) - t^{1-loop}(u;M) Q_2(u) = \nonumber\\
&&\!\!\!\!\!\!\!=\left(2-i\f{\gm^{(1-loop)}}{2} (u+i/2) \right) Q_1(u+i)+
\left(2+i\f{\gm^{(1-loop)}}{2} (u-i/2) \right) Q_1(u-i)+ \nonumber\\
&&\!\!\!+\,t^{2-loop}(u;M)\, Q_1(u)
\eqnx
where $\gm^{(1-loop)}(M)=8\, S_1(M)$ is the 1-loop energy and
\eq
t^{2-loop}(u;M)=-\left(4+\f{2M+1}{2}\gm^{(1-loop)}\right)
\eqx 
This is the only mild nonlinearity
in the dependence of the 2-loop Baxter equation on $Q_1(u)$. In \cite{KRZ}, a solution
of the above equation was found taking $Q_1(u)=\Qt_1(u)$. It is given explicitly
as
\eqn
\Qt_2(u) &=& \f{1}{2} \cdot 8 S_1(M) \, \f{\partial}{\partial \dl}\; {}_3F_2(-M,M+1+2\dl,1/2-iu; 1+\dl,1 | 1)_{| \dl=0} - \nonumber\\
&& - \f{\partial^2}{\partial \dl^2}\; {}_3F_2(-M,M+1,1/2-iu; 1+\dl,1-\dl | 1)_{| \dl=0} - \nonumber\\
&& - \f{\partial^2}{\partial \dl^2}\; {}_3F_2(-M,M+1,1/2-iu+\dl; 1+\dl,1+\dl | 1)_{| \dl=0}
\eqnx
Hence the nontrivial 2-loop part of $Q_2$ corresponding to the physical 1-loop solution (\ref{e.Qone})
is
\eq
\!\!\!\!\!\f{ \f{1}{2} i \sinh (2\pi u)}{\sin \pi M} \left[
\left(1\!-\!i\,\tan \f{\pi M}{2} \coth \pi u\right) \Qt_2(u) -
\left(1\!+\!i\,\tan \f{\pi M}{2} \coth \pi u\right) \Qt_2(-u) \right]
\eqx
We will denote it in what follows by $Q_2^{bare}(u)$ in order to emphasize that it is still not 
the final physical 2-loop solution.


Using the two series representations of hypergeometric functions given in appendix~\ref{s.Qtwo}
one can show that $Q_2^{bare}(u)$ has second order poles at $u=\pm i/2$.
We find explicitly at $u\sim i/2$:
\eq
Q_2^{bare}(u) \sim \f{2}{(u-i/2)^2} -\f{2i \left( 2 S_1(M) +\pi \tan \f{M\pi}{2} \right)}{u-i/2} + \oo{(u-i/2)^0}
\eqx
and at $u\sim -i/2$:
\eq
Q_2^{bare}(u) \sim \f{2}{(u+i/2)^2} +\f{2i \left( 2 S_1(M) +\pi \tan \f{M\pi}{2} \right)}{u+i/2} + \oo{(u+i/2)^0}
\eqx
The leading $2^{nd}$ order poles can be canceled by adding
\eq
Q_2^{bare}(u) \longrightarrow Q_2^{bare}(u) -2\pi^2\, \tanh^2 \pi u\, Q_1(u)
\eqx
Of course this modification can be absorbed into a $g$-dependent gauge factor.
However the remaining $1^{st}$ order poles have opposite residues at $u=i/2$ and
$u=-i/2$ so they cannot be canceled by a periodic function times $Q_1(u)$.
We can cancel them, however, by a term proportional to $\Qt_1(u)+\Qt_1(-u)$.
The asymptotic behaviour of this linear combination has a nonvanishing component proportional
to $u^M$ --- this is allowed by our proposal as other terms\footnote{Unfortunately
we lack reliable asymptotic estimates for $\Qt_2(u)$. We will therefore
examine the asymptotics of our final solution numerically.} are behaving
like $\sinh(2\pi u) \cdot u^{-1-M}$, so the $u^M$ component is exponentially suppressed
in accordance with our key proposal of section~\ref{s.key}.
 
The final form of the 2-loop solution $Q_2(u)$ is thus
\eq
Q_2(u)=Q_2^{bare}(u) \underbrace{-2\pi^2\, \tanh^2 \pi u\, Q_1(u) +
\f{\pi^2}{\cos^2 \f{M\pi}{2}} 
(\Qt_1(u)+\Qt_1(-u))}_{\text{necessary to cancel poles}} -c(M) Q_1(u)
\eqx
The final term is added in order to have $Q_2(i/2)=0$ which is convenient
for comparision with the formulas of \cite{KRZ,BBKZ} analytically continued
to arbitrary $M$. The explicit rather complicated expression for $c(M)$ is given
in a Mathematica notebook included in the arXiV submission \cite{nbnum}.

As a check of the 2-loop solution $Q_2(u)$ obtained above, we have numerically
evaluated the following expressions\footnote{Here we used the identities among harmonic sums given in appendix~\ref{s.analcont} in order to simplify these expressions before taking 
the analytical continuation to noninteger $M$. Moreover there is an overall numerical factor relative to \cite{BBKZ} coming from a different definition of $g$.} \cite{BBKZ} for the first and second derivatives
at $u=i/2$:
\eqn
Q_2(i/2) &=& 0 \qq \text{(normalization)}\\
Q'_2(i/2) &=& 8i (S_3-S_{-3}+S_1(S_2+S_{-2}) +2 S_{2,1}) \\
Q''_2(i/2) &=& 32 S_{-3} S_1+ 32 S_1^2 (S_2+S_{-2}) +32 S_1 S_3 +\\
 & & +64 S_{-3,1}+64 S_{-2,2}-128 S_{-2,1,1}
\eqnx
We found excellent agreement (up to a relative accuracy of at least $10^{-6}$) with the above expressions analytically continued
to $M=-9/10,\, 2/3$ and $33/10$ using the formulas of appendix~\ref{s.analcont}.
However, due to the derivatives w.r.t. parameters of the hypergeometric functions
appearing in $\Qt_2(u)$ and the nontrivial cancellation of poles exactly at $u=\pm i/2$, 
a precise numerical evaluation is somewhat involved. We give some details on that
in appendix~\ref{s.num} and attach a Mathematica notebook with that calculation
to the arXiV submission \cite{nbnum}.

We checked that for sample values of $M$, the behaviour of $Q_2(u)/\sinh(2\pi u)$ is
decreasing with $u$ (for $u$ up to around $10\sim 12$ and $M=7/3,\, 11/3$) in a way which is consistent
with the behaviour $1/u^{M+1} (1+c\, \log u)$, however the numerics seem to destabilise
for larger $u$ and we cannot reliably fit the exponent $M+1$. But certainly we may rule out the component 
proportional to $u^M$ in accordance with our proposal.

\section{Summary and outlook}

In this paper we identified asymptotic conditions for a solution of the Baxter equation
for twist-2 operators which works for any complex value of the spin. This
solution reproduces results such as energies and higher conserved charges which 
have been previously obtained in the conventional fashion
of first finding an analytical expression for the anomalous dimensions for
integer spin in terms of nested harmonic sums, and then performing an 
analytical continuation of these harmonic sums according to the procedure
outlined in~\cite{ANALCONT}.

The main interest of working directly for complex spin is that in this way we can bypass
the stage of finding an analytical expression for integer spin which becomes
prohibitively complicated at higher loop level (c.f. the results at 5 loops in
\cite{TWISTTWO5}) and virtually impossible in the current exact formulations
of the spectral problem at any coupling through TBA/FiNLIE/QSC. 

The identification of the asymptotic conditions for the Baxter function for any complex spin
may aid in constructing a formulation of TBA/FiNLIE/QSC which could
be used to study BFKL properties at any coupling. However, one has to note that
implementing these conditions in a numerical formulation for $M<-1/2$ may be
quite challenging as then, the $u^{M}$ branch becomes subleading at infinity.   

There are numerous open problems for further research. Firstly, it would be good
to understand the physical justification of the present proposal.
The conventional justifications in terms of a construction of a Baxter
operator for integrable spin chains \cite{korbaxt,matthiassl} would not neccessarily be
applicable here, as we even lack a direct hamiltonian construction
of the relevant system for complex~$M$. 

Secondly, it would be very interesting to understand the links with
the interrelations between the LO BFKL and anomalous dimensions explored in~\cite{Kor1,Kor2}.
These approaches seem to be very different from the present one, especially as 
some of their formulas become singular in the present setup. 

Last but not least, the outstanding open problem would be to implement these conditions 
as an ingredient of an exact TBA/FiNLIE/QSC formulation valid at any coupling.  

\bigskip

\noindent{\bf Acknowledgments.} I would like to thank Zoltan Bajnok for many discussions 
and Benjamin Basso for various comments. This work was supported by NCN grant 2012/06/A/ST2/00396.

\appendix

\section{Asymptotics of $\Qt_1(u)$}
\label{s.appQt}

Let us sketch how the asymptotic formula (\ref{e.qtiasympt}) can be motivated.
We start from the standard series representation of the 
hypergeometric function
\eq
\Qt_1(u)= \f{1}{\Gm(-M) \Gm(M+1) \Gm(1/2-iu)} \sum_{k=0}^\infty \f{\Gm(k-M) \Gm(k+M+1)
\Gm(1/2-i u+k)}{k!^3}
\eqx
We will now use the approximation valid at large $u$:
\eq
\label{e.gammaas}
\f{\Gm(1/2-i u+k)}{\Gm(1/2-iu)} \longrightarrow (-iu)^k
\eqx
and substitute it back into the series expression. The series can be summed up to get
\eq
{}_2 F_2(-M, M+1; 1, 1| -iu)
\eqx
Now performing a series expansion at $u=\infty$ we get
finally the formula
\eq
\label{e.qtiasrepeat}
\Qt_1(u) \sim e^{i\pi \f{M}{2}} \f{\Gamma(1+2M)}{\Gamma(1+M)^3} u^M \left(1+\ldots\right)
-i e^{-i\pi \f{M}{2}} \f{\Gamma(-1-2M)}{\Gamma(-M)^3} u^{-1-M} \left(1+\ldots\right)
\eqx
Here we ignored a term proportional to 
$e^{-i u}$ which is not seen either in a numerical check of (\ref{e.qtiasrepeat}) which we performed, nor in 
a related computation using a quite different approach in \cite{LipatovDV}.
This may be an artefact of the approximation (\ref{e.gammaas}) which is not entirely
reliable as subleading terms at higher orders in $k$ which are discarded
are of the same order as lower terms which are kept. We consider it more as heuristics
to obtain a formula which we subsequently verify numerically.

\section{Harmonic sums --- some identities and analytical continuation}
\label{s.analcont}

In order to reduce the complexity of finding the analytical continuation
of nested harmonic sums entering the formulas in \cite{KRZ,BBKZ} we used
several identities following from \cite{HSrelations}:
\eqns
S_{1,1} &=& \f{1}{2} S_1^2 +\f{1}{2} S_2 \\
S_{1,1,1} &=& \f{1}{6} S_1^3+ \f{1}{2} S_1 S_2+ \f{1}{3} S_3 \\
S_{1,-2} &=& S_1 S_{-2} +S_{-3}-S_{-2,1} \\
S_{1,-2,1} &=& -2 S_{-2,1,1}+S_{1} S_{-2,1}+S_{-3,1}+S_{-2,2} \\
S_{1,2,1} &=& -2 S_{2,1,1}+S_{1} S_{2,1}+S_{3,1}+S_{2,2} \\
S_{1,1,-2} &=& S_{-2,1,1}-S_{1} S_{-2,1}-S_{-3,1}-S_{-2,2}+S_{1}S_{-3}+S_{-4}+\f{1}{2} S_{-2}(S_{1}^2+S_{2}) \\
S_{1,1,2} &=& S_{2,1,1}-S_{1} S_{2,1}-S_{3,1}-S_{2,2}+S_{1}S_{3}+S_{4}+
\f{1}{2} S_{2}(S_{1}^2+S_{2}) \\
S_{1,1,1,1} &=& \f{1}{4} S_{4}+\f{1}{8} S_{2}^2+\f{1}{3} S_{1}S_{3}+
\f{1}{4} S_{1}^2 S_{2}+\f{1}{24} S_{1}^4 \\
S_{1,-3} &=& -S_{-3,1}+S_{-3}S_{1}+S_{-4} \\
S_{1,3} &=& -S_{3,1}+S_{1}S_{3}+S_{4} \\
S_{2,-2} &=& -S_{-2,2}+S_{2}S_{-2}+S_{-4} \\
S_{2,2} &=& \f{1}{2} (S_{2}^2+S_{4}) \\
S_{-2,-2} &=& \f{1}{2} (S_{-2}^2+S_{4})
\eqnsx

Below we quote formulas for the analytical continuation of the harmonic sums
which appear when evaluating conserved charges for the 1- and 2-loop case.
The parameter $a$ in the first two formulas is assumed to be positive.
\eqns
S_{a}(M) &=& \zeta(a) - \f{(-1)^a}{(a-1)!} \psi_{a-1}(M+1) \\
S_{-a}(M) &=& \zeta(a) (2^{1-a}-1) -\f{(-1)^a}{(a-1)!} \f{1}{2^a} \left(
\psi_{a-1}(1+M/2)-\psi_{a-1}((1+M)/2) \right) \\
S_{21}(M) &=& -\f{5}{8} \zeta(3)+ \sum_{m=0}^\infty  \f{(-1)^m}{(m+M+1)^2}
\left( \psi(m+M+2)-\psi(1) \right) \\
S_{-2,2}(M) &=& S_{-2,2}(\infty)- \sum_{l=0}^\infty \f{(-1)^{l+1}}{(l+m+1)^2} S_2(l+M+1) \\
S_{-3,1}(M) &=& S_{-3,1}(\infty)- \sum_{l=0}^\infty \f{(-1)^{l+1}}{(l+m+1)^3} S_1(l+M+1) \\
S_{1,1}(M) &=& \f{1}{2} \left( S_1(M)^2+S_2(M) \right) \\
S_{-2,1,1}(M) &=& S_{-2,1,1}(\infty) - \sum_{l=0}^\infty \f{(-1)^{l+1}}{(l+m+1)^2} S_{1,1}(l+M+1)
\eqnsx

\section{Properties of $\Qt_2(u)$}
\label{s.Qtwo}

In order to analyze the singularity structure of $\Qt_2(u)$ near $u=\pm i/2$
it is neccessary to perform expansions of a more general type of hypergeometric
function than the one appearing at 1-loop level, namely
\eq
{}_3F_2(-M,M+1+2\dl_1,1/2-iu+\dl_2;1+\dl_1+\dl_2+\dl_3,1+\dl_2-\dl_3|1)
\eqx
An expansion in the lower half plane can be obtained directly from the
standard definition of the ${}_3F_2$ similarly as (\ref{e.lower1}).
The key difficulty lies in generalizing the representation in the upper half plane
(\ref{e.upper1}). The formulas of~\cite{LipatovDV} using representations
of Legendre functions are not directly applicable in this case.

The idea of deriving such an expression is to start with the expression
\eqn
{}_3F_2(a,b,\f{1}{2}-iu+\bt;c,1+\al+\bt|1) &=&
\f{
\Gm(1+\al+\bt)}{\Gm(1/2+iu-\al) \Gm(1/2-iu+\bt)} \cdot\nonumber\\
&& \hspace{-2.5cm}\cdot \int_0^1 (1-z)^{iu-\f{1}{2}+\al} 
z^{-i u-\f{1}{2}+\bt} {}_2F_1(a,b;c|z)\, dz
\eqnx
which follows directly from power series definitions of the respective functions.
Then one expresses ${}_2F_1(a,b;c|z)$ in terms of ${}_2F_1(a,b;a+b-c+1|1-z)$ and
$(1-z)^{c-a-b} \cdot {}_2F_1(c-a,c-b;c-a-b+1|1-z)$, and integrates the power
series representations term by term. The final expression for 
${}_3F_2(a,b,\f{1}{2}-iu+\bt;c,1+\al+\bt |1)$
is as follows:
\eqns
&& \hspace{-1.0cm} {}_3F_2(a,b,\f{1}{2}-iu+\bt;c,1+\al+\bt|1) =\f{\pi \Gm(c) \Gm(1+\al+\bt)}{\sin \pi(c-a-b) \Gm(c-a) \Gm(c-b) \Gm(a) \Gm(b)} 
\bigg[ \nonumber\\
&&\sum_{k=0}^\infty \f{\Gm(a+k) \Gm(b+k)  \Gm(1/2+\al+k+i u)}{\Gm(a+b-c+1+k) \Gm(1+\al+\bt+k)
\Gm(1/2+\al+i u) k!} - \nonumber\\
&&\hspace{-0.4cm} - \sum_{k=0}^\infty
\f{\Gm(c-a+k) \Gm(c-b+k) \Gm(1/2-a+\al-b+c+k+i u)}{
\Gm(c-a-b+1+k) \Gm(1-a+\al-b+\bt+c+k) \Gm(1/2+\al+iu) k!} \bigg]
\eqnsx
The above expressions were also neccessary to derive the constant $c(M)$ appearing
in our final expression for $Q_2(u)$. Since the expression for $c(M)$ is rather
complicated, it is given in a Mathematica notebook attached to the arXiV submission \cite{nbnum}.

\section{Some details on the numerical evaluation of the 2-loop solution}
\label{s.num}

In order to perform numerical checks of the charges of the 2-loop solution we face two main
difficulties. Firstly, the elementary 2-loop solution $\Qt_2(u)$ is difficult to
evaluate numerically in Mathematica as it involves derivatives of ${}_3F_2$ w.r.t.
parameters of the hypergeometric function. Secondly, we need to evaluate derivatives
of $Q_2(u)$ at $u= i/2$, where the elementary 2-loop solution has $3^{rd}$ order poles.
Of course, these poles will get canceled by the overall factor of $\sinh 2\pi u$ and
subtractions of appropriate 1-loop solutions, but a precise numerical evaluation
is therefore difficult.

We perform the numerical evaluation in two steps. First we adjust the $c(M)$
coefficient in $Q_2(u)$ so that $Q_2(\pm i/2)=0$. This is done analytically using
series representations of $Q_2(u)$ derived in appendix~\ref{s.Qtwo}.
Then we construct a Chebyshev grid
of 20 points between $u=\pm i/2$. Once we will evalute the values of $Q_2$ at
the interior points of this grid, we will be able to evaluate very precisely
derivatives at $u=i/2$ using Chebyshev differentiation matrix.

It remains thus to evaluate the values of $\Qt_2(u)$ at a number of points
in the \emph{interior} of the interval $(-i/2,i/2)$. 

Now each term of $\Qt_2(u)$ is 
a derivative w.r.t. $\dl$ at $\dl=0$ of a hypergeometric function
e.g.
\eq
\f{\partial}{\partial \dl}\; {}_3F_2(-M,M+1+2\dl,1/2-iu; 1+\dl,1 | 1)_{| \dl=0}
\eqx
For each of these points $u \in (-i/2,i/2)$ we will use a second Chebyshev grid in $\dl$
of 21 points in the interval $(-1/2000,1/2000)$. We will use Mathematica to evaluate
the hypergeometric functions for these $\dl$ and use Chebyshev differentiation
to extract the derivative w.r.t $\dl$ at $\dl=0$. We attach a Mathematica notebook
with this computation to the arXiV submission \cite{nbnum}.


\begin{thebibliography}{99}

\bibitem{BFKL}   L.~N.~Lipatov,
  { ``Reggeization of the vector meson and the
  vacuum singularity in nonabelian gauge theories,''}
  Sov.\ J.\ Nucl.\ Phys.\  {\bf 23} (1976) 338
  [Yad.\ Fiz.\  {\bf 23} (1976) 642];\\
  E.~A.~Kuraev, L.~N.~Lipatov and V.~S.~Fadin,
  { ``The Pomeranchuk singularity in nonabelian gauge theories,''}
  Sov.\ Phys.\ JETP {\bf 45} (1977) 199
  [Zh.\ Eksp.\ Teor.\ Fiz.\  {\bf 72} (1977) 377];\\
  I.~I.~Balitsky and L.~N.~Lipatov,
  { ``The Pomeranchuk singularity in Quantum Chromodynamics,''}
  Sov.\ J.\ Nucl.\ Phys.\  {\bf 28} (1978) 822
  [Yad.\ Fiz.\  {\bf 28} (1978) 1597].

\bibitem{NLO}   A.~V.~Kotikov and L.~N.~Lipatov,
  ``DGLAP and BFKL equations in the N=4 supersymmetric gauge theory,''
  Nucl.\ Phys.\ B {\bf 661} (2003) 19
   [Erratum-ibid.\ B {\bf 685} (2004) 405]
  [hep-ph/0208220].

\bibitem{JP}   R.~A.~Janik and R.~B.~Peschanski,
  ``High-energy scattering and the AdS / CFT correspondence,''
  Nucl.\ Phys.\ B {\bf 565} (2000) 193
  [hep-th/9907177].


\bibitem{BPS}   R.~C.~Brower, J.~Polchinski, M.~J.~Strassler and C.~-ITan,
  ``The Pomeron and gauge/string duality,''
  JHEP {\bf 0712} (2007) 005
  [hep-th/0603115].


\bibitem{COSTA}   M.~S.~Costa, V.~Goncalves and J.~Penedones,
  ``Conformal Regge theory,''
  JHEP {\bf 1212} (2012) 091
  [arXiv:1209.4355 [hep-th]].

\bibitem{LipLast}   A.~V.~Kotikov and L.~N.~Lipatov,
  ``Pomeron in the N=4 supersymmetric gauge model at strong couplings,''
  Nucl.\ Phys.\ B {\bf 874} (2013) 889
  [arXiv:1301.0882 [hep-th]].

\bibitem{TOAPPEAR} R. Janik, P. Laskoś-Grabowski, to appear

\bibitem{KLRSV}   A.~V.~Kotikov, L.~N.~Lipatov, A.~Rej, M.~Staudacher and V.~N.~Velizhanin,
  ``Dressing and wrapping,''
  J.\ Stat.\ Mech.\  {\bf 0710} (2007) P10003
  [arXiv:0704.3586 [hep-th]].


\bibitem{TWISTTWO4}   Z.~Bajnok, R.~A.~Janik and T.~Lukowski,
  ``Four loop twist two, BFKL, wrapping and strings,''
  Nucl.\ Phys.\ B {\bf 816} (2009) 376
  [arXiv:0811.4448 [hep-th]].


\bibitem{TWISTTWO5}   T.~Lukowski, A.~Rej and V.~N.~Velizhanin,
  ``Five-Loop Anomalous Dimension of Twist-Two Operators,''
  Nucl.\ Phys.\ B {\bf 831} (2010) 105
  [arXiv:0912.1624 [hep-th]].

\bibitem{TBA}   N.~Gromov, V.~Kazakov, A.~Kozak, P.~Vieira,
  ``Exact Spectrum of Anomalous Dimensions of Planar $\mathcal{N} = 4$ Supersymmetric Yang-Mills Theory: TBA and excited states,''
  Lett.\ Math.\ Phys.\  {\bf 91} (2010) 265,
  [arxiv:0902.4458 [hep-th]];\\
  D.~Bombardelli, D.~Fioravanti, R.~Tateo,
  ``Thermodynamic Bethe Ansatz for planar AdS/CFT: A Proposal,''
  J.\ Phys.\ A: Math.\ Theor.\ {\bf 42} (2009) 375401,
  [arxiv:0902.3930 [hep-th]];\\
  G.~Arutyunov, S.~Frolov,
  ``Thermodynamic Bethe Ansatz for the $AdS_5\times S^5$ Mirror Model,''
  JHEP {\bf 0905} (2009) 068,
  [arxiv:0903.0141 [hep-th]].


\bibitem{FINLIE}   N.~Gromov, V.~Kazakov, S.~Leurent and D.~Volin,
  ``Solving the AdS/CFT Y-system,''
  JHEP {\bf 1207} (2012) 023
  [arXiv:1110.0562 [hep-th]]


\bibitem{BHNLIE}   J.~Balog and A.~Hegedus,
  ``Hybrid-NLIE for the AdS/CFT spectral problem,''
  JHEP {\bf 1208} (2012) 022
  [arXiv:1202.3244 [hep-th]].

\bibitem{QSC}   N.~Gromov, V.~Kazakov, S.~Leurent and D.~Volin,
  ``Quantum spectral curve for $AdS_5/CFT_4$,''
  arXiv:1305.1939 [hep-th].


\bibitem{ANALCONT}   A.~V.~Kotikov and V.~N.~Velizhanin,
  ``Analytic continuation of the Mellin moments of deep inelastic structure functions,''
  hep-ph/0501274.


\bibitem{Jaroszewicz}  T.~Jaroszewicz,
 ``Gluonic Regge Singularities And Anomalous Dimensions In QCD,'' Phys.\
 Lett.\  B {\bf 116} (1982) 291.


\bibitem{Basso}   B.~Basso,
  ``Scaling dimensions at small spin in N=4 SYM theory,''
  arXiv:1205.0054 [hep-th].

\bibitem{KRZ}   A.~V.~Kotikov, A.~Rej and S.~Zieme,
  ``Analytic three-loop Solutions for N=4 SYM Twist Operators,''
  Nucl.\ Phys.\ B {\bf 813} (2009) 460
  [arXiv:0810.0691 [hep-th]].

\bibitem{BBKZ}   M.~Beccaria, A.~V.~Belitsky, A.~V.~Kotikov and S.~Zieme,
  ``Analytic solution of the multiloop Baxter equation,''
  Nucl.\ Phys.\ B {\bf 827} (2010) 565
  [arXiv:0908.0520 [hep-th]].

\bibitem{HSrelations}   J.~Blumlein and S.~Kurth,
  ``Harmonic sums and Mellin transforms up to two loop order,''
  Phys.\ Rev.\ D {\bf 60} (1999) 014018
  [hep-ph/9810241].

\bibitem{G}   N.~Gromov,
  ``On the Derivation of the Exact Slope Function,''
  JHEP {\bf 1302} (2013) 055
  [arXiv:1205.0018 [hep-th]].


\bibitem{KorQt}   S.~E.~Derkachov, G.~P.~Korchemsky and A.~N.~Manashov,
  ``Evolution equations for quark gluon distributions in multicolor QCD and open spin chains,''
  Nucl.\ Phys.\ B {\bf 566} (2000) 203
  [hep-ph/9909539].

\bibitem{LipatovDV}   H.~J.~De Vega and L.~N.~Lipatov,
  ``Interaction of reggeized gluons in the Baxter-Sklyanin representation,''
  Phys.\ Rev.\ D {\bf 64} (2001) 114019
  [hep-ph/0107225].


\bibitem{GKpriv} N. Gromov, V. Kazakov, private communication. See also talk by N. Gromov at IGST 2013.

\bibitem{nbnum} {\it Mathematica} notebook in the source file of the arXiv submission.

\bibitem{korbaxt}   S.~E.~Derkachov, G.~P.~Korchemsky and A.~N.~Manashov,
  ``Separation of variables for the quantum SL(2,R) spin chain,''
  JHEP {\bf 0307} (2003) 047
  [hep-th/0210216].

\bibitem{matthiassl}   R.~Frassek, T.~Lukowski, C.~Meneghelli and M.~Staudacher,
  ``Baxter Operators and Hamiltonians for 'nearly all' Integrable Closed gl(n) Spin Chains,''
  arXiv:1112.3600 [math-ph].

\bibitem{Kor1}   G.~P.~Korchemsky, J.~Kotanski and A.~N.~Manashov,
  ``Multi-reggeon compound states and resummed anomalous dimensions in QCD,''
  Phys.\ Lett.\ B {\bf 583} (2004) 121
  [hep-ph/0306250].

\bibitem{Kor2}   S.~E.~Derkachov, G.~P.~Korchemsky, J.~Kotanski and A.~N.~Manashov,
  ``Noncompact Heisenberg spin magnets from high-energy QCD. 2. Quantization conditions and energy spectrum,''
  Nucl.\ Phys.\ B {\bf 645} (2002) 237
  [hep-th/0204124].

\end{thebibliography}
\end{document}